\documentclass{aa}            

\usepackage{graphicx}
\usepackage{txfonts}

\begin{document}

\title{NGC 4051: Black
hole mass and  photon index-mass accretion rate correlation 
}

\author{  Elena Seifina\inst{1} 
\and
Alexandre Chekhtman\inst{2}         
\and
Lev Titarchuk \inst{3}         
          }

   \institute{LAPTh, Universite Savoie Mont Blanc, CNRS, B.P. 110, Annecy-le-Vieux
F-74941, France, \
Lomonosov Moscow State University/Sternberg Astronomical Institute,
Universitetsky
Prospect 13, Moscow, 119992, Russia, \email{seif@sai.msu.ru}
        \and
George Mason University, College of Science, 4400 University Drive, 
Fairfax, VA 22030 
\email{achekhtm@gmu.edu};
        \and
    Dipartimento di Fisica, University di Ferrara, Via Saragat 1,
I-44122, Ferrara, Italy,  \email{titarchuk@fe.infn.it}         }

   \date{Received 
        ;       accepted 
}

\abstract
{ 
We present a discovery of the correlation between the X-ray spectral (photon)  index 
and  mass accretion rate  observed  in an active galactic nucleus, NGC 4051.  
We analyzed spectral  transition episodes observed in NGC 4051   using XMM/{\it Newton}, 
{\it Suzaku} and {\it RXTE}.  We applied a scaling technique for a black hole (BH) mass 
evaluation which uses  a  correlation between the photon index and 
normalization of the    seed (disk)  component, which is proportional to a mass accretion rate. 
We developed an analytical model that shows the spectral (photon) index of the BH emergent 
spectrum  undergoes an evolution from  lower to higher values depending  on 
 a mass accretion rate in the accretion disk.
We considered Cygnus X-1 and GRO~J1550-564 as  
reference sources for which distances, inclination  angles and the BH masses are  
evaluated  by  dynamical measurements.  
Application of the scaling technique for the photon  index$-$mass 
accretion rate correlation provides an estimate of the black hole mass in NGC 4051 to 
be more than $6\times10^5$ solar masses.
} 


\titlerunning{On evaluation of BH mass in NGC 4051}

\maketitle

\section{Introduction}

Determination of masses of Galactic black holes (BHs) is one of the most important tasks in modern astronomy. There are two methods for BH mass determination. One relies on 
determination of the mass function using optical observations. While the theoretical mass function is a function of the two masses and the inclination angle, the value of the observationally inferred mass function (the minimum possible mass of the compact
object) depends only on the velocity half-amplitude K and
the orbital period $P$ and is independent of the inclination angle $f (M)\propto PK^3$.
\cite{oro03} summarized the measurements of the rotational
velocities and inclinations for 17 BH binaries.

However, this mass function method cannot be applied to active galactic nuclei (AGN) and for this case the second method of BH mass estimate  using  the photon index-mass accretion rate  correlation [see \cite{st09}, hereafter ST09] is potentially  more promising because it applies only information on X-ray spectral  characteristics of  a compact object.  It is worthwhile to point out that 
an evolution of the photon  index for super-massive BH sources (AGN) was not systematically analyzed while the presence of fast evolution of X-ray spectral shape similar to one observed in a number of galactic BHs was reported also for AGNs [see \cite{vup11}]. In this paper we present a detailed analysis of X-ray spectral evolution in the AGN NGC 4051  and its application  for  a BH mass estimate.

%
%

\begin{figure*}[ptbptbptb]
\centering
\includegraphics[scale=0.7,angle=0]{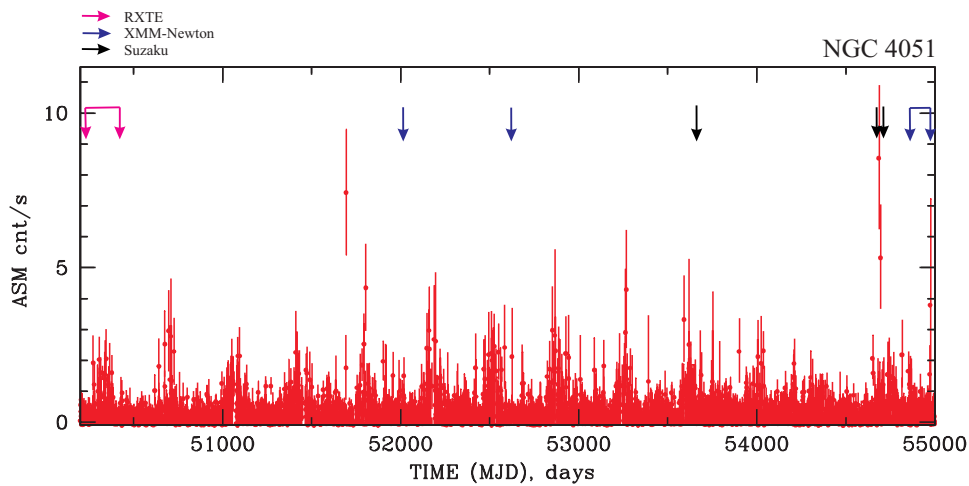}
\caption{Evolution of ASM/{\it RXTE} count rate during 1996 -- 2009 observations of NGC~4051.
Vertical arrows (at the top of panel) indicate temporal distribution of the {\it RXTE} (pink), XMM-Newton (blue) and Suzaku (black) data sets listed in Tables \ref{tab:list_XMM}$-$\ref{tab:list_RXTE}.
 }
\label{RXTE evol}
\end{figure*}
%
%
\begin{figure*}[ptbptbptb]
\centering
\includegraphics[scale=1.0,angle=0]{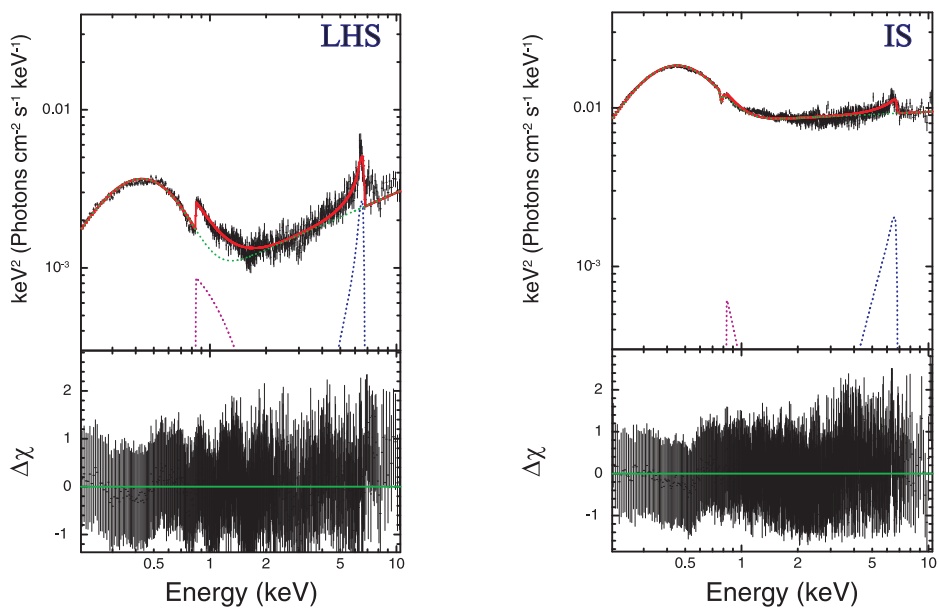}
\caption{Representative EF(E) spectral diagrams for the low-hard   state  (left panel)
and  for the intermediate state   (right panel).
Data are taken from the XMM-$Newton$ observations 0157560101 (left panel, X2 data set),
and 0109141401 (right panel, X1 data set).
The data are shown by black crosses and the spectral model components are displayed
by dashed green,  purple and blue  lines for the BMC, Laor, and redge components, respectively.
 }
\label{2_sp_state_NGC}
\end{figure*}

%
%

\begin{figure*}[ptbptbptb]
\centering
\includegraphics[scale=0.7,angle=0]{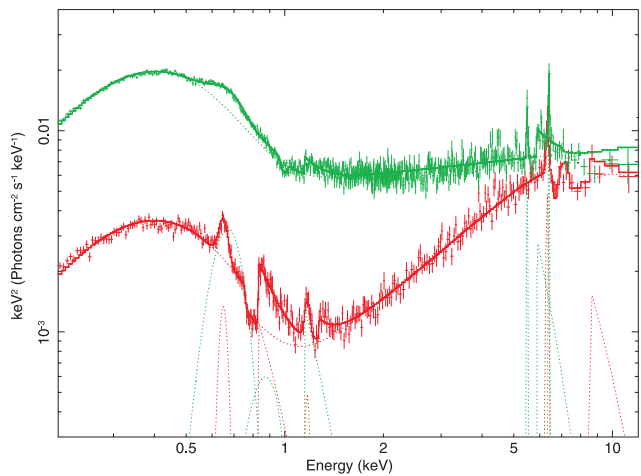}
\caption{Spectral   evolution  using  XMM-{\it Newton} data  of NGC4051 for  different spectral states: the LHS spectrum (May 11, 2009 -red points) and
 the IS  spectrum (May 15, 2009 - green points). 
Curves are related to the best-fit model (see details in text). }
\label{XMM_spectra}
\end{figure*}

%
%

\begin{figure*}[ptbptbptb]
\centering
\includegraphics[scale=0.7,angle=0]{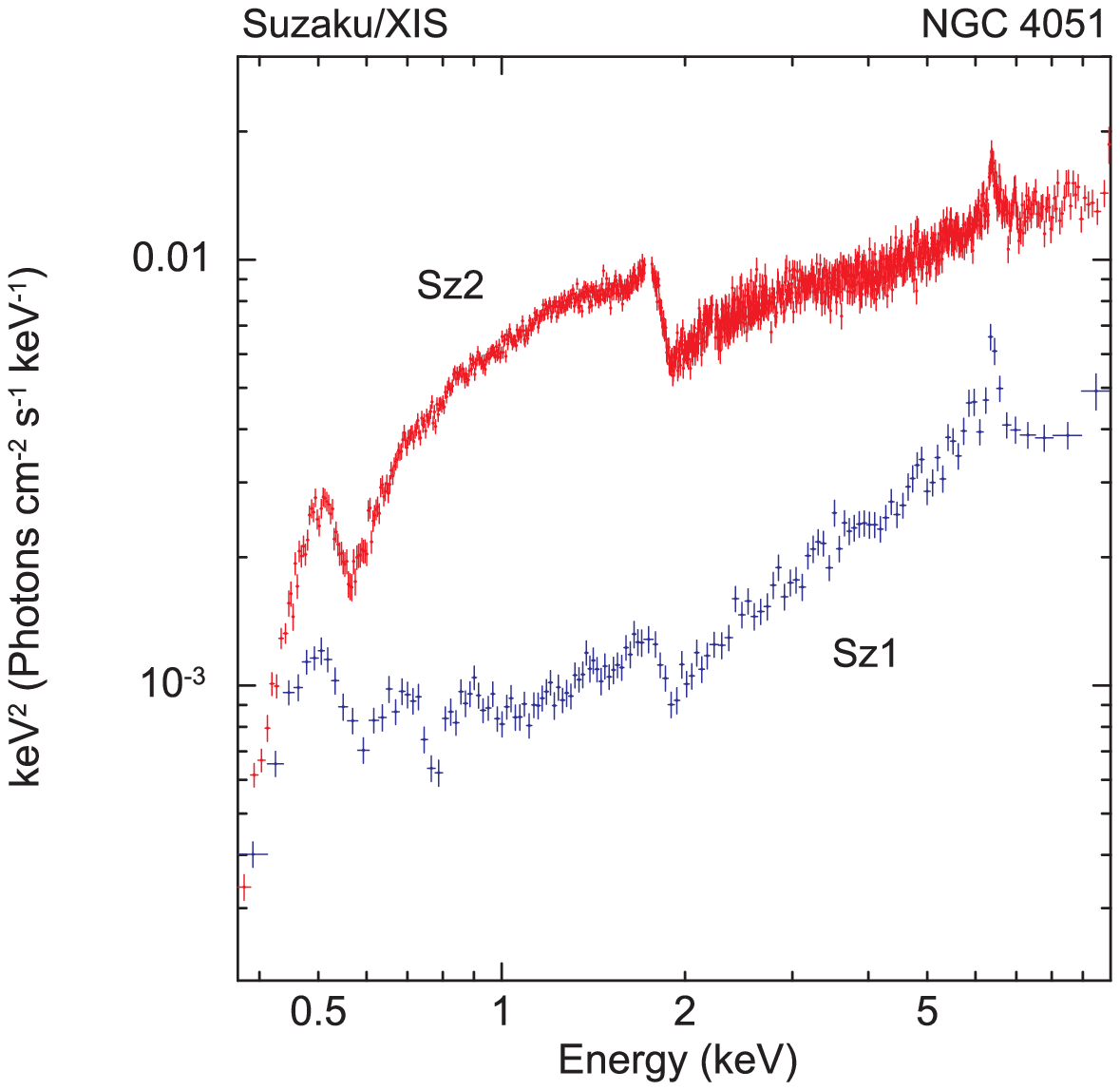}
\caption{Two $E F_E$ spectral diagrams during the LHS (2005 Nov, blue) and the IS 
[2008 Nov, 6 -- 12   (red)]. Data taken from 
$Suzaku$ observations 700004010 (Sz1, low-hard) and 703023010 (Sz2, intermediate state).
}
\label{suzaku_spectra}
\end{figure*}

 Observational appearances of BHs in Galactic sources  is conventionally described in terms of the BH spectral  state classification
\citep[see][ for different flavors of BH states definitions]{rm,bell05,kw08}.
A general  BH state classification for 
five major BH states:{ quiescent}, { low-hard} (LHS), 
{ intermediate} (IS), {high-soft} (HSS) and { very high} 
states (VHS) is accepted by the community.  When a BH transient  goes into outburst 
it leaves the quiescent state and enters the LHS, a low luminosity
state with the energy spectrum dominated by the thermal Comptonization component 
related to  a  weak thermal component. The photon spectrum in the LHS  
is presumably a result of the  Comptonization (upscattering)
of soft  photons, that originated  in a relatively weak  inner part of the accretion disk,  off 
electrons of  the hot ambient plasma [see e.g., \cite{st80}].
Variability in  the LHS is high (fractional root-mean-square variability
is up to 40\%) and  presented by a flat-top broken power law (white-red noise) shape, accompanied by quasi-periodic oscillations (QPOs) in the range of 0.01-30 Hz, observed as narrow peaks in the power  density spectrum (PDS).
 The HSS photon spectrum is characterized by a prominent thermal 
component which is probably a signature of  a strong emission coming from a geometrically thin accretion disk.
A weak power-law component is also present at a level  
of not more than 20\% of the total source flux. In the HSS the flat-top variability 
ceases, QPOs disappear, and the PDS acquires a pure power-law shape.
The total  variability in the HSS is usually about 5\% fractional rms.
The IS is a transitional stage between the LHS and the HSS.

We have concentrated our efforts on a study of  correlations between the photon index 
and the  accretion  disk (soft photon)  luminosity. This  correlation 
pattern carries the most direct information on a BH mass and  a source  distance.
We also investigated a possibility that the shape of this correlation
pattern can provide a direct signature of the bulk motion (converging) flow onto  a BH, 
or in other words, the BH signature [see \cite{tz98} and \cite{lt99} for more details on this subject].

The authors of ST09 argue that that  the saturation effect seen in the index-mass accretion rate correlation  is a signature of a converging  flow, when the mass accretion rate, $\dot M$   exceeds the Eddington limit, which can only  exist in BH sources.  We note that in the neutron star (NS) sources  the mass accretion rate is, in most cases, 
less than the Eddington rate limit. 
As far as we know,  in only Sco X-1 does it reaches the Eddington limit [compare
\cite{st09}; \cite{ts09}; \cite{st10}; \cite{st11};  \cite{st12}; \cite{stf13}; \cite{tsf13} and \cite{sts14}].

In the observed  LHS and  IS  a small fraction of the disk emission component is  seen directly. 
The Compton cloud almost fully covers  the inner part of the disk.
The energy spectrum is dominated by a Comptonized component seen as 
a power law at energies higher than the characteristic energy of the disk blackbody (BB) emission.
 To calculate the   normalization of the seed (disk) BB component we modeled the spectrum
using a generic Comptonization model  (BMC)  (see Titarchuk et al. 1997) which consistently convolves 
a seed  BB  emission with the Green's  (response) function of the Compton corona  (as a broken power law) to
produce the Comptonized component (see details in \S 2). 
Calculated in this way the spectrum normalization is directly related to a BH mass, 
source distance and geometry.  This allows us to apply the BB 
normalization as a scaling variable and to impose a
constraint on a BH mass and distance ratios. 
Thus, we used the  index-disk normalization 
correlation to constrain a  BH mass for  NGC 4051.
  As our reference values for the scaling measurements, we used the 
previously measured  BH masses, source distances and inclinations
for Galactic sources, Cygnus X-1 and GRO J1550-564. 
We also used a previous estimate of a BH mass for NGC 4051 obtained by Denney et al. (2009),  (1.73$\pm$0.55)x10$^6$ solar masses  in order to compare with our evaluation of it.

The descriptions of XMM/{\it Newton}, {\it Suzaku} and {\it RXTE}  data  are given in \S \ref{data}.
We provide the details of the spectral state transitions analysis in  \S \ref{transitions}.  We  describe a  scaling method technique   and  present the results of interpretation  of the data using   the scaling method  in  \S 4. 
We discuss  our results and 
emphasize on a BH mass estimate  in NGC 4051 in \S 4. We  make  our final conclusions in \S  5 

\section{Observations and data reduction \label{data}}

For the study presented in this paper we have analyzed data for spectral transitions in NGC 4051  
observed using   XMM/{\it Newton}, {\it Suzaku} and {\it RXTE}. We  analyzed 
a set of  data taken during  years when the source showed a transition from  the LHS to  the HSS. 
We used the archival  XMM/{\it Newton}, 
$Suzaku$ and {\it RXTE} data from the HEASARC\footnote{http://heasarc.gsfc.nasa.gov/}. For  XMM/{\it Newton} there are 15 observations, 40 ks each collected in May-June 2009,  two   50 ks observations collected on May, 16 2001 and on November, 22 2002 (see Table  \ref{tab:list_XMM}). Data reduction was done using SAS software version 10.0.0. Four {\it Chandra} observations had much smaller observational time and thus, larger 
statistical errors and were 
used mainly to confirm the XMM/{\it Newton} results by an independent detector with different systematic errors.
 XMM/{\it Newton} and {\it Chandra} data have been extracted and analyzed in the 0.2-12 keV energy band.

We studied NGC~4051 using also the {\it Suzaku} data,  for November 10 of 2005 as well as for November 6 and 23 of the 
2008 observations (see Table~\ref{tab:list_Suzaku}). We also used  {\tt HEASOFT software package} (version 6.13) and 
calibration database  (CALDB) released on  February 10,   2012  for  XIS. 
Since the background is dominant in the lower energy band, we used photons in the 1 -- 10 keV (for XIS0, and 3) 
and 1 -- 7 keV (for XIS 1) energy bands. 
The data reduction and spectral analysis are performed following 
  the {\it Suzaku} Data Reduction Guide\footnote{http://heasarc.gsfc.nasa.gov/docs/suzaku/analysis/}. 
 We obtained cleaned event files by re-running the {\it Suzaku} {\tt pipeline} implementing the
latest calibration database (CALDB) available since  January 20 of  2013,    and also applied the associated screening criteria files. 
We  extracted our spectra 
from the cleaned event files using XSELEC and we generated  responses  for each
detector utilizing  the XISRESP script with a medium resolution.
The spectral and response files for the front-illuminated
detectors (XIS0, 1 and 3) were combined using the FTOOL
ADDASCASPEC, after confirmation of  their consistency. 
In addition, we grouped the spectra in order to have a minimum of 20 counts per energy bin. 



We analyzed available data using the public archive \citep{bradt93} using the {\it RXTE} observations made from May, 1996 to May, 2001. In Figure \ref{RXTE evol} we show the light curve evolution   for the   {\it RXTE} 1996-2009 observations of NGC 4051. 
Standard tasks of the LHEASOFT/FTOOLS 6.21 software package were applied for data processing. For spectral analysis we 
used PCA {\it Standard 2} mode data, collected in the 3 -- 30~keV energy range, using 
PCA response calibration (ftool pcarmf v11.7).
The standard dead time correction procedure was applied to the data. 
We  subtracted a background corrected  in  off-source observations. 
Systematic error of 0.5\% was applied to all analyzed {\it RXTE}  spectra. 
In  Table~\ref{tab:list_RXTE} we provide the necessary information  on the {\it RXTE} observations. 
We modeled  all 
energy spectra 
using the XSPEC astrophysical fitting software. 

To fit  XMM/{\it Newton},  and {\it Suzaku} 
 ($BMC$) component, the Laor line component  with the energies   $\sim6.5$ keV, which are presumably due to 
the iron emission line (see Fig. \ref{2_sp_state_NGC}) and the recombination edge (redge) component modified by gabs absorption. 
All the {\it RXTE} spectra were fitted using our  composite model consist of the BMC and Laor line components.

%
%
\begin{figure*}[ptbptbptb]
\centering
\includegraphics[scale=0.8,angle=0]{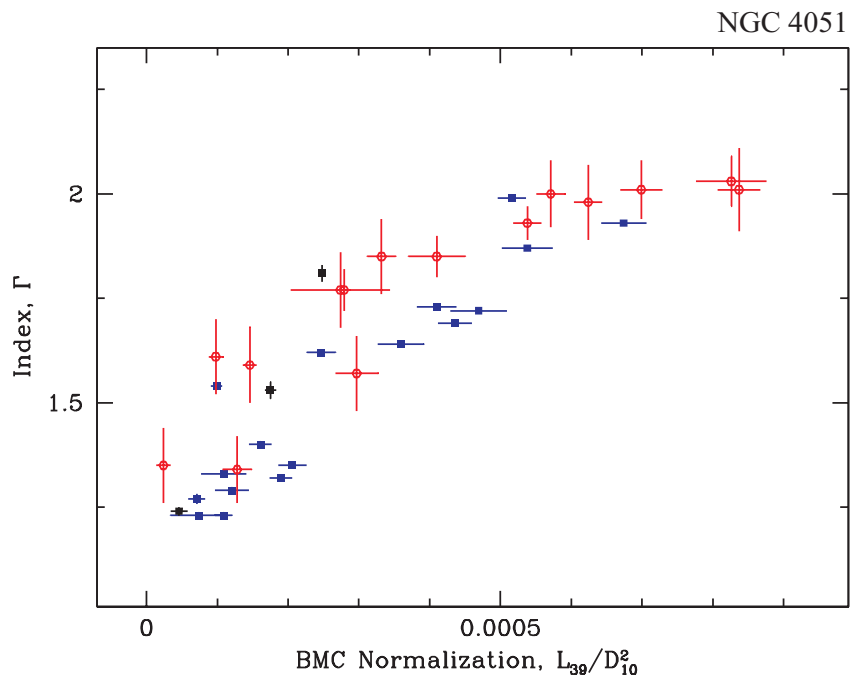}
\caption{Photon index versus BMC normalization (proportional to mass accretion rate) for  outburst transitions (1996 -- 2009) in
NGC~4051. Red triangles, and blue and black squares correspond to {\it RXTE}, XMM-$Newton,$ and
$Suzaku$ observations, respectively.
 }
\label{index__mdot_cor_for_NGC4051}
\end{figure*}

%
%

\begin{figure*}[ptbptbptb]
\centering
\includegraphics[scale=0.8,angle=0]{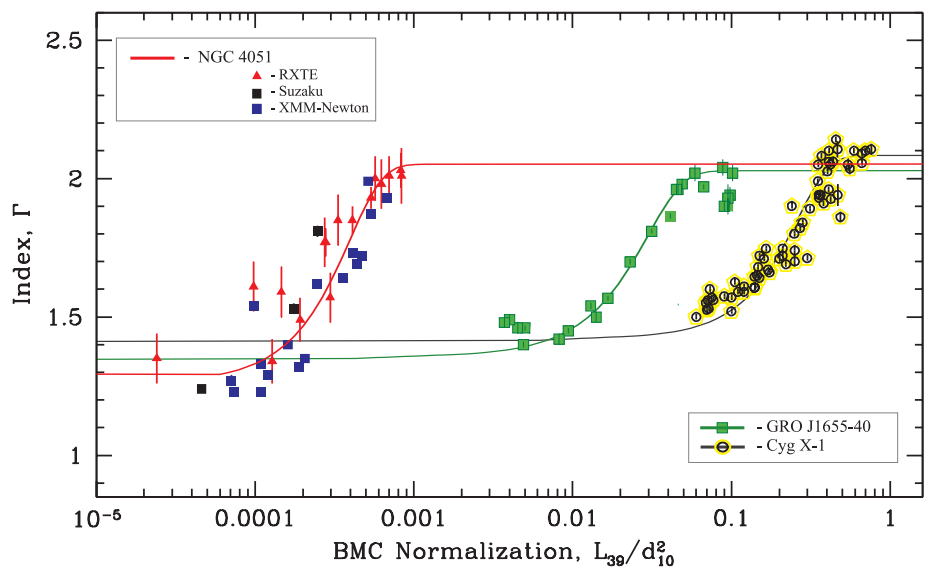}
\caption{Scaling of the photon index versus normalization for NGC~4051 (red line -- target
source), and GRO~J1655--40 and Cygnus~X--1 (reference sources, green and black lines, respectively). Red triangles, black and blue squares correspond to
{\it RXTE}, $Suzaku$ and XMM-$Newton$ data. Green squares and
yellow-black circles correspond to {\it RXTE} data for GRO~J1655--40 and Cygnus~X--1,
respectively.
 }
\label{mass_determination_for_NGC4051_chandra_xmm}
\end{figure*}

In particular,  the { BMC} model describes the outgoing spectrum as a convolution 
of the  { seed} BB spectrum where  the model parameters are
 normalization $N_{bmc}$,  a color temperature $kT_{col}$,
 using  the Comptonization Green's  function.
Similarly to the ordinary BB XSPEC model
the normalization $C_{N}$ is a ratio of source luminosity
to square of the distance d
\begin{equation}
C_{N}=\biggl(\frac{L}{10^{39}\mathrm{erg/s}}\biggr)\biggl(\frac{10\,\mathrm{kpc}}{d}\biggr)^2.
\end{equation}  
The resulting  model spectrum is also characterized by a parameter $\log(A)$ related
to a Comptonized (illumination) fraction,  $f=A/(1+A)$ and  the Green's  function spectral index
$\alpha=\Gamma-1$, where $\Gamma$ is the  photon index. 

The two reasons for using the BMC model are firstly, the BMC, by the nature is  a model  which is applicable to a general case when
soft (disk) photons gain energy not only due to the thermal Comptonization  but also due to
a dynamical or bulk motion Comptonization. Secondly, the BMC  has  normalization of the { seed}  (disk) photons $C_{N}$, which  is 
 proportional to the mass accretion rate in the disk.  

The adopted spectral model well describes  the most data sets used in our analysis. 
A value of the reduced 
$\chi^2_{red}=\chi^2/N_{dof}$, where $N_{dof}$ is  a number of degrees of freedom
for a given fit, is always  around 1.0 for all of XMM, 
{\it Suzaku} and {\it RXTE}  observations (see Tables \ref{tab:analysis_XMM}-\ref{tab:analysis_RXTE}).
 
We incorporate the Laor-line profile to fit the line component. The line feature has a statistical
significance of  (3 -- 10) $\sigma$ depending on the spectral states. This line  is variable 
and exhibits equivalent width (EW) in the range of 50 -- 700 eV across the data , see Tables \ref{tab:analysis_XMM}-
\ref{tab:analysis_RXTE}. We found that an addition of  the 
Laor-line component significantly improves the fit quality of the LHS and IS  spectra.

We revealed absorption edges in the   fits of the NGC~4051 spectra, 
which can be interpreted as an evidence of a clumpy, multi-temperature outflow around NGC~4051, particularly in the IS.
We also applied the recombination edge emission model, redge\footnote{https://heasarc.gsfc.nasa.gov/xanadu/xspec/XspecManual.pdf} 
for all $Suzaku$ and XMM-$Newton$ spectra, which considerably improves the fit quality of the LHS spectra. 
The parameters of the redge model are: $E_c$ threshold energy; $T_{pl}$  plasma temperature (keV), and $N_{redge}$,  normalization 
in photons/cm$^2$/s in the line.

Thus, we identified a complete set of 
spectral transitions in a period when the source state  changes from the  LHS to the  IS 
(or vice versa). We should establish that NGC 4051  does not reach  the real  HSS when the photon 
index of X-ray spectrum usually  exceeds two (see e.g.,  ST09).  The photon index saturation  value is also determined by the electron temperature of the converging flow (see for example, Laurent \& Titarchuk 1999, hereafter LT99).
In Tables \ref{tab:analysis_XMM} -- \ref{tab:analysis_RXTE} 
we present the details of the model  fits to the spectral  data
for each of the observations related to a paricular spectral  state defined by a value of the spectral (or photon) index. 

As seen above we used all available XMM-{\it Newton} and Suzaku archived observations of NGC4051, which are distributed accordingly to the specific observation schedule.  We also used all particular well-monitored {\it RXTE} observations of NGC 4051 where we observe the spectral changes, particularly changes of the spectral (or photon) indices. In these cases we established the correlation of the photon index, $\Gamma$ vs normalization of the photon component  which is identical  to normalization of the soft (disk) component. 

\section{Correlations between the spectral (photon) index and mass accretion rate  during spectral 
 state transitions \label{transitions}}

In Figure~\ref{2_sp_state_NGC} we show two representative EF(E) spectral diagrams for the 
LHS and the IS states   (red lines). Data were taken from XMM-$Newton$ observations 0157560101 
(left panel, ``X2'' data set, LHS), and 0109141401 (right panel, ``X1'' data set,
IS). 

In Figure \ref{XMM_spectra}  we demonstrate a spectral evolution   
using XMM/{\it Newton}  data for the LHS (red curve) and IS (green curve) states (2001,  2009). 
 It is worth pointing out  a strength of the low energy component at   $\sim 0.5$ keV  in  the IS  and its weakness in  the LHS. 
 The  XMM-$Newton$ spectra, which are characterized by the highest energy resolution quality, 
  clear demonstrate a number of additional emission and absorption features in the LHS 
(see Fig.~\ref{XMM_spectra}, red line). 
However, these features are not well pronounced in the   IS  spectra  (e.g., Figure~\ref{XMM_spectra}, green line). 
Thus,  in this Figure we show the complete set of the spectra, which were detected 
during the spectral  evolution. 

We should emphasize  the presence  of the Laor line component and the recombination edge emission component, redge near 6.4 keV in all XMM/$Newton$ spectra (see Figs. \ref{2_sp_state_NGC}-\ref{XMM_spectra}).   For all spectra we also applied a Gaussian absorption line model, gabs using  as  parameters, the line energy,  $E_{line}$, (in keV)  the line width, $\sigma$ (in keV) and the line optical  depth at the line center.
Our model for the data fits is   $gabs x (BMC + Laor + Redge)$.

In Figure~\ref{suzaku_spectra} we also show  the spectra of NGC~4051 for the  2005 and 2008 outburst event epoches with another 
X--ray mission, $Suzaku$. The data cover two spectral states of the source, in our classification: the  LHS (blue), the IS (red). 

The normalization of the spectra  (proportional to the mass accretion rate)   positively correlates
with the photon index. The correlation pattern for NGC 4051 presented in Figure \ref{index__mdot_cor_for_NGC4051}
demonstrate  a clear saturation of the photon index which 
is  also identified in several Galactic and extragalactic BH sources [see ST09,  
\cite{st11};  \cite{st12}; \cite{stf13}; \cite{tsf13}, Seifina et al. (2014), (2017); Titarchuk \& Seifina (2016a),(2016b), (2017)].

On the other hand  \cite{ft11} (hereafter FT11) and later Seifina \& Titarchuk (2011), (2012), (2013); Seifina et al. (2013), (2015), (2016); Titarchuk  et al. (2013), (2014),  (2015), (2016) 
show that the index stays the same  in  a  number of NS sources independently of a mass accretion rate.  One sees almost no change of the index when a particular NS source evolves  from the low to high  luminosity states. Thus, even phenomenologically the index  evolves for BHs and stays  around two for NSs. ST09 and FT11 present the strong theoretical  arguments for such particular behaviors of the index in BH and NS sources, respectively.  
   
The index saturation level  for the spectral transition in NGC 4051   has values close to two. It can be  due to the fact that  
at lower accretion rates in the disk 
the cooling of X-ray emission  area [Compton cloud (CC)] by the disk photons is less 
efficient than that at high disk mass accretion rate $\dot M_d$. 
As a result, the electron  temperature $kT_e$ of the converging flow (CF) is still relatively  high (10$-$15 keV, see LT99) even when the total mass accretion $\dot M_{sub}+ \dot M_d$ rate is greater than 1.  As a result this temperature effect leads to more efficient upscattering in CF  and harder spectra. The direct connection between the electron  temperature of the 
CC and the hardness of the emergent spectrum were shown by \cite{lt99}, (2011) using Monte-Carlo simulations.
This type of the index-mass accretion rate correlation is also seen in  Cygnus X-1 and in a number of other  BHCs (see details in ST09). 

As one can see that our definition of the spectral state is not arbitrary.  Indeed, normalization of the soft component in the LHS is one order of magnitude  less than that in the IS (see Figs. \ref{2_sp_state_NGC}-\ref{suzaku_spectra}). 
The photon indices are also  different; at about 1.5 in the LHS and  2 in the IS  (see Figs. \ref{2_sp_state_NGC}-\ref{mass_determination_for_NGC4051_chandra_xmm}).

\subsection{Evolution of the spectral Index from the low-hard state to softer states in BHs: Analytical explanation}

We then estimated a value of the spectral index in the low-hard state and its evolution to softer states.
For this estimate we used the energy balance in the transition layer (TL)  dictated by Coulomb collisions of  protons (gravitational energy release), while inverse Compton 
and free-free emission are the main cooling channels [see a formulation of this problem in the pioneering work by \cite{zs69} 
and a similar consideration in \cite{bisn80}].   
For the characteristic electron temperature (3 keV $\la kT_e\la$ 50 keV)  of these regions, 
Compton cooling dominates over free-free emission and   
the relation between the energy flux  per unit surface area of the corona  $Q_{cor}$,  the radiation energy  density 
 $\varepsilon(\tau)$  and electron temperature ${\cal T}_e(\tau)$  can be  given by [see \cite{zs69} and \cite{tlm98}, hereafter TLM98]
\begin{equation}
Q_{cor} =  20.2\int^{\tau_0}_0 \varepsilon(\tau){\cal T}_e(\tau)d\tau,
\label{energy_balance}
\end{equation}
where  $\tau$ is  the Thomson optical depth in
the transition layer (TL). 
The distribution $\varepsilon(\tau)$ is obtained as a solution of the diffusion equation
\begin{equation}
\frac{d^2\varepsilon}{d\tau^2} =-q
\label{diffusion_equation}
\end{equation}
where $q$ is  a distribution of the sources in the corona due to the gravitational energy release of the accreted matter there and illumination of the corona by the soft flux of the disk. 
The parameter $q$ can be presented as 
\begin{equation}
q(\tau)=\frac{3Q_{tot}}{c}
\frac{1}{\tau_{0}}
\label{q_tau},
\end{equation}
where  $Q_{tot}=Q_{cor} + Q_{disk}$ is the sum of the corona (TL) and intercepted disk fluxes, respectively. 
 Combination of Eq. (\ref{diffusion_equation}) with two boundary conditions at the outer  and inner TL boundaries
\begin{equation}
\frac{\partial \varepsilon}{\partial \tau}-\frac{3}{2}\varepsilon |_{\tau=0} =\frac{\partial \varepsilon}{\partial \tau}+\frac{3}{2}\varepsilon|_{\tau=\tau_0}=0
\label{inner_outer_bound_conds}
\end{equation} 
  leads us  to the formulation of the TL boundary problem.

 The solution for $\varepsilon(\tau)$ is  given by
\begin{equation}
 \varepsilon(\tau)=\frac{Q_{tot}}{c} \left[1+ \frac{3}{2}\tau_0\left(\frac{\tau}{\tau_0} -
  \frac{\tau^2}{\tau_0^2}\right)\right].
\label{ene_vs_tau}
\end{equation}
We then rewrite the right-hand of Eq. (\ref{energy_balance}) as
\begin{equation} 
T_e=\frac{\int_0^{\tau_0}\varepsilon(\tau){\cal T}_e(\tau)d\tau}{\int_0^{\tau_0}\varepsilon(\tau)d\tau} 
\label{int_T_e_epsilon}
\end{equation} 
using the mean value theorem, where $T_e$ is the mean electron temperature.

The  following integral  of $ \varepsilon(\tau)$ [see Eq. ({\ref{ene_vs_tau}})] is
\begin{equation}
\int^{\tau_0}_0 \varepsilon(\tau) d\tau 
=\frac{Q_{tot}}{c}\tau_0(1+\tau_0/4).
\label{average_ene}
\end{equation}

If we substitute Formula  (\ref{average_ene}) into Eq. (\ref{energy_balance}), after some 
straightforward algebra we obtain
\begin{equation}
\frac{kT_e \tau_0 [1+\tau_0/4]}{m_e c^2}=
\frac{0.25}{1+Q_{disk}/Q_{cor}}.
\label{ktetau}
\end{equation}

To evaluate how the spectral (photon) index changes with ratio $Q_{disk}/Q_{cor}$ we use a  formula  for  the spectral index $\alpha$ 
\begin{equation}
\alpha=-\frac{3}{2}+\sqrt{\frac{9}{4}+\gamma}, 
\label{alpha_general}
\end{equation}
where $\gamma=\beta/\Theta$ and $\beta-$parameter, $\Theta \equiv kT_e/m_e c^2$ 
defined  in \cite{tl95}.
If we replace $\beta$ by  its diffusion limit  $\beta_{\rm diff}$ 
\begin{equation}
\beta_{\rm diff}=\frac{\pi^2}{3(\tau_0+4/3)^2}\approx\frac{\pi^2}{3\tau_0^2}\
\label{beta_diff}
\end{equation} 
for $\tau_0\gg 4/3$  using 
 we obtain  the diffusion spectral index as 
\begin{equation}
\alpha_{diff}= -\frac{3}{2}+\sqrt{\frac{9}{4}+ \gamma_{diff}}
\label{alpha_diff}
\end{equation}
where $\gamma_{diff}=\beta_{diff}/\Theta=(\pi^2/3\tau_0^2)/\Theta$.
Using this formula  for $\gamma_{diff}$ and  Eq. (\ref{ktetau})
we obtain  that 
\begin{equation}
\gamma_{diff}\approx \frac{\pi^2}{3}(1+Q_{disk}/Q_{cor})
\label{gamma_diffusion}
\end{equation}

We can also rewrite Eq. (\ref{alpha_diff}) for $\alpha_{diff}$ in a  more convenient   form 
\begin{equation}
\alpha_{diff}= \frac{\gamma_{diff}}{3/2+\sqrt{9/4+ \gamma_{diff}}}
\label{alpha_diff_mod}
\end{equation}
from which the following two  asymptotic behaviors of $\alpha_{diff}$  are obvious:
 When $ Q_{disk}/Q_{cor}\ll 1$  it follows that $\gamma_{diff}\sim 3$  and then  
\begin{equation}
\alpha_{diff}\sim 0.7.
\label{small_alpha_diffusion}
\end{equation}While for   $Q_{disk}/Q_{cor}\gg 1$  (or with a rise of the mass accretion rate) $\gamma_{diff}$ increases which leads to an increase of $\alpha_{diff}$:  
\begin{equation}
\alpha_{diff}=\gamma^{1/2}_{diff}\gg1. 
\label{great_alpha_diffusion}
\end{equation}Thus, using formulas ({\ref{gamma_diffusion}$-$\ref{great_alpha_diffusion}) one can see that 
$\alpha_{diff}$  rises when 
$Q_{disk}/Q_{cor}$ increases.

The  index evolution   including its saturation with
the  mass accretion rate were analyzed observationally and numerically by ST09 and \cite{lt11} (hereafter LT11). 
Here we  estimate analytically that  the spectral index $\alpha$  
really rises from $\sim0.7$ when    $Q_{disk}/Q_{cor}\ll1$  and then the emergent spectrum becomes softer when the illumination of the innermost part of the accretion flow  by the soft (disk) photons $Q_{disk}$ increases (see e.g., Fig. \ref{index__mdot_cor_for_NGC4051}).

Indeed, in Figures  \ref{index__mdot_cor_for_NGC4051}-\ref{mass_determination_for_NGC4051_chandra_xmm}  we demonstrate  that the best-fit photon index really grows with a mass accretion rate, $\dot m$ (or with $L_{39}/d_{10}^2$, see below
 Eq. \ref{lumin_m}).  But when $\dot m$ increases the disk soft photon  illumination of the transition layer,  $Q_{disk}$ increases too and thus, formulas (\ref{alpha_diff_mod}$-$\ref{great_alpha_diffusion}) qualitatively  describes the index rise with a mass accretion rate.
The behavior of the photon index versus a mass accretion rate in the converging flow was studied in detail by LT99 and LT11 with taking into account all effects of General Relativity.  
Here, we describe the index increase qualitatively with the mass accretion rate.
    
\section{Discussion
\label{discussion}} 

The accretion process onto a BH is far from
being completely understood. Theoretical models which are available for
astronomers to explain observational phenomena usually deal
with only narrow aspects of a broad observational picture.
However, in recent years
a concept of  the transitional layer (TL) in the accretion flow has 
emerged. In the framework of this model several major observational
aspects of accreting BHs find the natural explanation.
We argue that this concept is a strong candidate to provide a basis
for a theory of accretion process onto compact objects
which would account for the current observational picture of BH sources.
Moreover, this paradigm may provide us with a first direct BH signature 
observationally confirmed in multiple sources.

The starting point in the development of the TL
paradigm is the notion of the necessary deviation
of the rotational profile in the innermost part of the disk 
set by the conditions at (or near) a compact object for which rotation is presumed to be much slower than  the Keplerian frequency near the object. This problem was first considered   by \cite{tlm98} who  showed  that the adjustment of the Keplerian rotation in  the accretion disk to the sub-Keplerian rotation of the central  object  leads to 
the formation  of the inner hot Compton corona
 
Currently, BH identification in X-ray observational astronomy is based solely
on a mass of compact object.
Namely, a compact X-ray source is classified as a BH if
it is well established that its mass exceeds the upper mass limit of a
stable rotating neutron star, namely 3.2 M$_{\odot}$ 
\citep[see e.g.,][] {rr74}.
To the date  there is only one  widely accepted  method for mass determination related to the 
measurement of the mass function $f(M)$ based on optical spectroscopy (see above). But this method the 
mass determination used for binaries cannot be applicable for AGNs. 
 
We have evaluated  a BH mass in NGC 4051 using scaling relationships in correlations between  the index and mass accretion rate observed during the state transitions in Galactic and extragalactic accreting BHs.  We applied a previously proposed technique for BH mass determination using  scaling the index$-$X-ray spectrum normalization (see ST09) which contains information on  source distance and BH mass. 

\subsection{The scaling technique and BH mass evaluation in NGC 4051 \label{analysis}}

One of the goals of this study is to apply X-ray observational data 
for NGC 4051 to infer its fundamental characteristics. To do this
we used the BH spectral state paradigm combined with the scaling
laws expected to be observed in spectral and timing data during state
transitions. 

The  scaling law, which we have used as a basis for our mass  determination analysis,
relates a source flux $F$ detected by an observer on the Earth, a disk flux  $L$ and distance $d$, namely
\begin{equation}
\frac{F_r}{F_t}=\frac{L_r}{L_t}\, \frac{d^2_t}{d^2_r} .
\end{equation}
where subscripts $r$ and $t$
denote reference and target sources respectively, and  $F$ stands for the source flux detected by an  observer on the Earth. 
The disk flux $L$ can be represented as 
\begin{equation}
L=\frac{GM_{bh} \dot M}{R_*}=\eta(r_*) \dot m  L_{\rm Edd}.  
\label{lumin}
\end{equation}
Here $R_{*}=r_{*}R_{\rm S}$ is an effective radius at which the main energy release takes place, 
$R_{\rm S}=2GM/c^2$ is the Schwarzschild radius, $\eta=1/(2r_*)$, $\dot m=\dot M/\dot M_{crit}$ is dimensionless mass accretion rate in units of the critical mass accretion rate
$\dot M_{crit}=L_{\rm Edd}/c^2$ and $L_{\rm Edd}$ is the Eddington luminosity.

On the other hand 
\begin{equation}
L_{\rm Edd}=\frac{4\pi GMm_pc}{\sigma_{\rm T}}
\label{ed_lumin}
\end{equation}
i.e., $L_{Edd}\propto M$ and thus, using Eqs. (\ref{lumin}-\ref{ed_lumin})
we have 
\begin{equation}
L\propto\eta(r_*) \dot m m.
\label{lumin_m}
\end{equation}

We assume  that both $\dot m$ and $\eta$ are  the same for two different 
sources in the same spectral state, which leads to $L_r/L_t=M_r/M_t$. 
In our spectral analysis we determined the normalization of  the { seed} photon 
radiation, which is supplied by an accretion flow (disk) prior to 
Comptonization. The ratio of these normalizations for
target and reference sources in the same spectral state can be written as
\begin{equation}
s_N=\frac{N_r}{N_t}=\frac{L_r}{L_t}\frac{d^2_t}{d^2_r}f_G.
\label{scaling_normalization}
\end{equation}
Here, $f_G$ is a  geometrical factor which  takes into account the difference in viewing angles
of the seed photon emission area (the inner disk) by the Earth observer for these given  sources.
Therefore, in the case of radiation coming directly from the disk it has
the value $f_G=(\cos \theta)_r/(\cos \theta)_t$, where $ \theta$ is the inclination
angle of the inner disk. This formula for $f_G$, however, may not be  accurate
 because the geometry  of both the inner disk and the corona
 can be different from the plane (disk) geometry. In BH states of interest (the LHS and the IS)  
  they can be described, for instance, 
 by quasi-spherical configuration. Despite this uncertainty in the determination of  $f_G$, we adopted the above form $f_G$ in which  $\theta\sim i$ if the information on the system inclination 
angle $i$ is available.  
In doing this we introduced an additional assumption
that the inner accretion disk and the system inclination angles are the same. 
This assumption, however, proved to be in good agreement with this type of analysis.

Using Eq. (\ref{scaling_normalization})  and a relation  $L_r/L_t=M_r/M_t$ in hand, the task of
BH mass measurement  for a target source is reduced to 
the determination of scaling coefficients  $s_N$ with
respect to the data for a reference source. The appropriate technique 
has already been implemented in a number of BH sources [see  ST09, Titarchuk \& Seifina (2016a), (2016b), (2017) hereafter TS16a, TS16b, TS17]. 
 
\subsection{Estimate of a black hole mass}
To estimate a BH mass, $M_{BH}$ of NGC~4051, we chose Galactic
sources [GRO~J1655--40 
and Cyg~X--1 (see ST09)] as the reference sources 
for  which BH masses 
and  distances 
were well established now (see Table~\ref{tab:par_scal}).   
For  a BH mass estimate of NGC~4051 we  used the BMC normalizations, $N_{BMC}$ of these reference sources.  

Thus, we scaled  the index vs  $N_{BMC}$  correlations for these reference sources  with that of 
the target source NGC~4051 (see Fig.~\ref{mass_determination_for_NGC4051_chandra_xmm}). 
The value of the  index saturation in NGC~4051, $\Gamma_{sat}^{ngc}\sim 2.05$,  is close to 
$\Gamma_{sat}^{gro}\sim 2.03$ of GRO~J1655--40 and to $\Gamma_{sat}^{cyg}\sim 2.09$ of Cyg~X--1. 
The scaling procedure was implemented in a similar way as in ST09, TS16a, TS16b.
 We introduced an analytical 
approximation  
of the $\Gamma(N_{bmc})$ correlation, 
fitted by a formula 
\begin{equation}
{\cal F}(x)= {\cal A} - ({\cal D}\cdot {\cal B})\ln\{\exp[(1.0 - (x/x_{tr})^{\beta})/{\cal D}] + 1\}
\label{scaling function}
\end{equation}
with $x=N_{bmc}$. Fitting  of the  observed correlation  by  this function ${\cal F}(x)$
provide us a set of the best-fit parameters $\cal A$, $\cal B$, $\cal D$, $x_{tr}$, and $\beta$.  
(see Table \ref{tab:parametrization_scal}).
More detailed description of these parameters is given in TS16a.

To implement this BH mass estimate for the target source one should rely on the same shape of the 
$\Gamma-N_{bmc}$ correlations for the target source and those for the reference sources.  To estimate a BH mass, 
$M_t$  of NGC~4051 (target source) one should slide the reference source correlation 
along $N_{bmc}-$axis  to that of the target source (see Fig.~\ref{mass_determination_for_NGC4051_chandra_xmm}).
\begin{equation}
M_t=M_r \frac{N_t}{N_r}
\left(\frac{d_t}{d_r}
\right)^2 f_G.
\label{scaling coefficient}
\end{equation}

In Figure~\ref{mass_determination_for_NGC4051_chandra_xmm} we demonstrate   the $\Gamma-N_{bmc}$ correlation  for NGC~4051 
using XMM-$Newton$ (blue squares), $Suzaku$ (black squares) and {\it RXTE} (red triangles) spectra (see Tables~\ref{tab:analysis_XMM} -- 
\ref{tab:analysis_RXTE}) along with those for the Galactic reference sources, Cyg~X--1 (black circles) and GRO~1655--40 (green squares). 
BH masses and distances for this target-reference pair are presented  in Table~\ref{tab:par_scal}. 

We used values of $M_r$, $M_t$, $d_r$, $d_t$, and $\cos (i)$ from Table~\ref{tab:par_scal} 
and then  calculated the lowest limit of the mass, using the best fit value of  $N_t= (8.1\pm 0.1)\times 10^{-3}$ 
taken them at the beginning of the index saturation  (see Fig. \ref{mass_determination_for_NGC4051_chandra_xmm}) and measured
in units of $L_{39}/d^2_{10}$ erg s$^{-1}$ kpc$^{-2}$ [see Table \ref{tab:parametrization_scal}
 for values of the parameters of function ${\cal F}(N_t)$ (see Eq. \ref{scaling function})].
Finally,  we obtained that $M_{ngc}\ga 6.1\times 10^5~M_{\odot}$ ($M_{ngc}=M_t$) 
assuming $d_{ngc}\sim$9.8 Mpc and  $f_G\sim1$. To determine the distance to NGC~4051 we used the formula

\begin{equation}
d_{ngc}=z_{ngc}c/H_0
\label{bllac_distance}
\end{equation} 
where  the redshift $z_{ngc}=0.00234$  for NGC~4051 (see { Wright 2006}), 
$c$ is the speed of light and $H_0=70.8\pm 1.6$ km s$^{-1}$ Mpc$^{-1}$  is the Hubble constant.  
We summarize these results in Table~\ref{tab:par_scal}.

Thus, we constructed 
the  scaling which allowed us to obtain  a BH mass of  $\ga6\times10^5$ solar masses  in NGC 4051 which is  our final evaluation  of BH mass using the  scaling method.
The obtained  BH mass estimate is in agreement with estimates  of  a variability time in optical H$_{\beta}$ line 
($M_{ngc}\sim 1.1\times 10^6 M_{\odot}$, Peterson  et al., 2000). Furthermore, the lower limit of the BH mass for NGC 4051 from 
reverberation mapping method ($6.13\times 10^5$ M$_{\odot}$, Kaspi et al. 2000) 
 is close to  that found in this paper.  We also refer to the BH mass estimate in NGC 4051,
$(1.7\pm 0.5)\times 10^6 M_{\odot}$ obtained by Denney et al. (2009) and this BH mass interval  indicates that our  estimate   does not contradict  to  the Denney's et al.  evaluation.

 Our scaling method was effectively applied  to evaluate BH masses of Galactic 
(e.g., ST09, STS13) and extragalactic black holes (TS16a,b; Sobolewska \&
Papadakis 2009; Giacche et al. 2014). Recently the scaling method was also  successfully implemented  to estimate BH masses 
of  two ultraluminous X-ray (ULX) sources M101 ULX--1 (TS16a) and ESO~243--49 HLX--1 (TS16b). These findings suggest  BH masses of  order of $10^4$ solar masses in these unique objects. 

We then applied  the scaling method  and found  the BH mass  in another extragalactic source NGC 4051. 
Thus,  the scaling technique  provides  a promising prospect for a BH mass evaluation  for  Galactic and extragalactic BHs,   
 TDE, and ULX sources.

\section{Conclusions \label{summary}} 

We performed a comprehensive study of a highly representative 
set of well observed spectral transitions in NGC 4051.
The goal of the study is to  explore the possibility 
of measuring the BH fundamental properties applying  X-ray 
data and search for a BH signature  in  NGC 4051 . 
We used the correlation between spectral and mass accretion rate 
properties during the state transitions as a main tool to estimate
 BH mass  in NGC 4051. 

We established  the spectral state transition  in this source
collecting X-ray  data from the {\it Newton} XMM, 
$Suzaku$ and  {\it RXTE} missions. We examined the correlation between the photon 
index of the Comptonized spectral component  and its normalization.  Analyzing  the behavior of the correlation patterns we utilize three basic scaling laws: the observed disk flux  is  proportional  to a BH mass,  the mass accretion rate and  inversely proportional  to  a square  of  distance to a particular source. We established that scalable correlation patterns do indeed contain information on 
BH mass.

Then we  combined  the correlation patterns 
in the  normalization domain  for Galactic X-ray BH binaries,
Cygnus X-1, GRO 1655-40,  for which BH masses and distances are known, and scaled them with that  
for NGC 4051. 
As a result we found   that all of these correlations are   self-similar   for these three BHs. 
The application of the scaling technique 
for the high precision measurements of BH masses requires { well sampled observations of
the source evolution through the outbursts and careful consideration of scalability 
of correlation patterns}.
 
 We show  that  the photon index of the BH emergent spectra of NGC 4051,
$\Gamma$ undergoes an evolution from  small to high values depending  on a mass accretion rate in  the source. 
Using this correlation along with scaling technique  gives us an  estimate of a BH mass in  NGC 4051,  $M_{bh}\ga 6\times 10^5 M_{\odot}$.
We also note that the referee of this paper also emphasizes that NGC 4051 is one of the  NLSy1s (narrow line Seyfert 1s), which are thought to have BHs with relatively low mass, of the order of $10^6$ solar masses. This is why the Earth observer   could even see the disk emission in X-rays (and is thus highly relevant to our model) 
(see Figs. \ref{2_sp_state_NGC}$-$\ref{XMM_spectra}).

\begin{acknowledgements}

E.S. acknowledges Pascal Chardonnet for useful discussions. L.T.  thanks Filippo Frontera for  wide discussion of the results of this paper and the referee for her/his interesting 
remarks on the content of the paper.
\end{acknowledgements}

\appendix
%
%

\begin{table*} 
    \caption{The XMM-{\it Newton} observations of NGC~4051 in the 0.2 -- 10 keV  range
}
    \centering
    \begin{tabular}{l l l l l c c}
      \hline\hline
Set number  & Obs. ID& Start time (UT)  &  End time (UT) & MJD interval & Mean count rate \\
  &        &                  &                &              &  (cts/s)     \\ 
 \hline                                   
X1 & 0109141401$^{(1,2)}$ & 2001 May 16 22:52:05 & 2001 May 17 20:22:31 & 52045.9 -- 52046.8 & 33.00$\pm$0.02\\
X2 & 0157560101$^{(2)}$ & 2002 Nov 22 06:09:51 & 2002 Nov 22 20:01:48 & 52600.2 -- 52600.8 & 6.21$\pm$0.01\\
X3 & 0606320101$^{(3)}$ & 2009 May 3  10:22:29 & 2009 May 3  23:03:02 & 54954.4 -- 54954.9 & 6.21$\pm$0.01\\
X4 & 0606320201$^{(3)}$ & 2009 May 5  10:15:55 & 2009 May 5  22:41:39 & 54956.4 -- 54956.9 & 12.81$\pm$0.02\\
X5 & 0606320301$^{(3)}$ & 2009 May 9  10:01:34 & 2009 May 9  18:49:22 & 54960.4 -- 54960.7 & 24.79$\pm$0.04\\
X6 & 0606320401$^{(3)}$ & 2009 May 11 10:00:35 & 2009 May 11 18:00:24 & 54962.4 -- 54962.7 & 6.18$\pm$0.02\\
X7 & 0606321301$^{(3)}$ & 2009 May 15 13:16:11 & 2009 May 15 21:38:31 & 54966.5 -- 54966.9 & 31.15$\pm$0.04\\
X8 & 0606321401$^{(3)}$ & 2009 May 17 09:41:17 & 2009 May 17 20:35:00 & 54968.4 -- 54968.8 & 19.31$\pm$0.03\\
X9 & 0606321501$^{(3)}$ & 2009 May 19 09:34:48 & 2009 May 19 09:28:43 & 54970.3 -- 54970.4 & 19.41$\pm$0.03\\
X10 & 0606321601$^{(3)}$ & 2009 May 21 09:27:52 & 2009 May 21 20:59:21 & 54972.3 -- 54972.8 & 38.95$\pm$0.04\\
X11 & 0606321701$^{(3)}$ & 2009 May 27 10:50:16 & 2009 May 27 21:29:10 & 54978.4 -- 54978.8 & 8.13$\pm$0.02\\
X12 & 0606321801$^{(3)}$ & 2009 May 29 09:15:27 & 2009 May 29 20:20:52 & 54980.3 -- 54980.8 & 11.55$\pm$0.03\\
X13 & 0606321901$^{(3)}$ & 2009 June 2 11:04:29 & 2009 June 2 21:11:35 & 54984.4 -- 54984.8 & 6.51$\pm$0.02\\
X14 & 0606322001$^{(3)}$ & 2009 June 4 10:41:05 & 2009 June 4 20:56:32 & 54986.4 -- 54986.8 & 11.46$\pm$0.03\\
X15 & 0606322101$^{(3)}$ & 2009 June 8 08:40:24 & 2009 June 8 19:08:04 & 54990.3 -- 54990.7 & 3.81$\pm$0.02\\
X16 & 0606322201$^{(3)}$ & 2009 June 10 08:21:42 & 2009 June 10 19:47:00 & 54992.3 -- 54992.8 & 10.34$\pm$0.02\\
X17 & 0606322301$^{(3)}$ & 2009 June 16 08:29:06 & 2009 June 16 20:13:36 & 54998.3 -- 54998.8 & 10.35$\pm$0.02\\
      \hline
      \end{tabular}
   \label{tab:list_XMM}
\tablebib{
(1) M$^c$Hardy et al. (2004); 
(2) \cite{haba08}; 
(3) Pounds\& King (2013).
}
\end{table*}

%
%

\begin{table*}
    \caption{
{Suzaku} observations of NGC~4051}
    \centering
    \begin{tabular}{lccccc}
      \hline \hline
Set number & Obs. ID& Start time (UT)  &  End time (UT) & MJD interval & Mean count rate \\
      \hline
Sz1 ................ & 700004010$^{(1,2)}$ & 2005 Nov 10 19:14:14 & 2005 Nov 13 10:21:24 & 53684.1 -- 53687.0 & 0.456$\pm$0.002\\
Sz2 ................ & 703023010$^{(1)}$ & 2008 Nov 6  07:39:03 & 2008 Nov 12 02:40:14 & 54776.6 -- 54782.1 & 2.331$\pm$0.003\\
Sz3 ................ & 703023020$^{(1)}$ & 2008 Nov 23 16:48:00 & 2008 Nov 25 13:43:43 & 54793.5 -- 54795.4 & 1.531$\pm$0.004\\
      \hline
      \end{tabular}
   \label{tab:list_Suzaku}
\tablebib{
(1) \cite{Lobban11}; 
(2) \cite{Terashima09}. 
}
\end{table*}

%

\begin{table*}
    \caption{{\it RXTE} observations of NGC~4051 in the 3 -- 20 keV  range   
}
        \centering
    \begin{tabular}{l l l c l c c}
      \hline
      \hline      
Observational ID& Start time (UT)  & MJD & Exposure \\
       &                  &        & (sec)     \\
      \hline
10301-01-04-00$^{(1)}$ & 1996-05-20 10:25:37.3 & 50223.4 &  943 \\
10301-01-07-00$^{(1)}$ & 1996-05-21 22:13:33.3 & 50224.9 &  1610 \\
10301-01-08-00$^{(1)}$ & 1996-05-22 04:06:53.3 & 50225.1 &  1210 \\
10301-01-09-00$^{(1)}$ & 1996-05-23 03:13:26.1 & 50226.1 &  1323 \\
10301-01-22-00$^{(1)}$ & 1996-05-29 07:18:46.1 & 50232.3 &  1045 \\
10301-01-58-00$^{(1)}$ & 1996-10-19 19:16:43.4 & 50375.8 &  1232 \\%
10301-01-60-00$^{(1)}$ & 1996-10-21 19:17:42.1 & 50377.8 &  1292 \\ %
10301-01-61-00$^{(1)}$ & 1996-10-22 13:20:24.6 & 50378.5 &  1026 \\ %
20318-01-01-00$^{(1)}$ & 1996-12-13 10:16:22.7 & 50430.4 &  9250 \\ %
20319-01-01-00$^{(1)}$ & 1996-11-10 17:37:05.4 & 50397.7 &  945  \\ %
20319-01-03-00$^{(1)}$ & 1996-12-10 19:19:56.1 & 50427.8 &  1018  \\ %
50153-01-01-01$^{(1)}$ & 2001-05-16 05:41:39.3 & 52045.2 &  18401 \\ %
50153-01-01-02$^{(1)}$ & 2001-05-16 15:19:39.4 & 52045.6 &  1777  \\ %
50153-01-01-03$^{(1)}$ & 2001-05-16 18:05:00.7 & 52045.7 &  12537 \\ %
50153-01-01-05$^{(1)}$ & 2001-05-16 03:49:03.6 & 52045.1 &  4595 \\
50153-01-01-06$^{(1)}$ & 2001-05-16 16:35:13.6 & 52045.6 &  2711 \\
      \hline
      \end{tabular}
   \label{tab:list_RXTE}
\tablebib{
(1) M$^c$Hardy et al. (2004) 
}
\end{table*}

%
%

\begin{table*}
    \caption{Best-fit parameters of spectral analysis for 2001, 2002 and 2009 with XMM/$Newton$ observations of 
NGC~4051 in 0.2 -- 10 keV energy range$^a$}
    \centering
    \begin{tabular}{l c c c c c c c c c c}
      \hline
      \hline
Observational& $kT_s$  &$\alpha=$  & $\log(A)$ & $N_{bmc}^b$ & $E_{line}$ & $N_{line}^b$ & $E_{c}$  & $kT_{pl}$ &  $N_{redge}^b$ &$\chi^2_{red}$ \\
ID           & (keV)    & $\Gamma-1$  &        &  (keV)      &            &              &   (keV)   &  (keV)  &    & (dof)         \\
\hline
0109141401 & 0.10(1) & 0.99(4) & -0.16(1) & 5.16(5) & 6.42(8) & 2.8(1) & 0.78(3) & 0.29(8) & 2.1(1) & 0.95 (1976) \\
0157560101 & 0.11(3) & 0.54(1) & -0.34(2) & 0.99(2) & 6.59(7) & 0.9(3) & 0.74(5) & 0.26(1) & 2.7(3) & 0.76 (1976) \\
0606320101 & 0.09(1) & 0.62(2) & -0.49(2) & 2.47(5) & 6.46(6) & 1.1(2) & 0.75(4) & 0.30(5) & 1.3(2) & 0.87 (1976) \\
0606320201 & 0.10(1) & 0.73(2) & -0.54(3) & 4.10(7) & 6.59(8) & 2.7(1) & 0.76(3) & 0.32(1) & 1.4(1) & 0.69 (1976) \\
0606320301 & 0.10(2) & 0.72(2) & -0.49(1) & 4.7(1)  & 6.51(4) & 2.5(4) & 0.77(2) & 0.35(4) & 1.6(3) & 0.96 (1976) \\
0606320401 & 0.10(1) & 0.23(1) & -0.14(6) & 0.74(8) & 6.59(7) & 0.7(3) & 0.78(7) & 0.34(2) & 0.9(4) & 0.72 (2157) \\
0606321301 & 0.10(3) & 0.87(3) & -0.51(2) & 5.38(9) & 6.41(5) & 3.2(2) & 0.74(5) & 0.33(1) & 2.3(2) & 0.85 (2157) \\
0606321401 & 0.10(2) & 0.69(4) & -0.54(4) & 4.36(6) & 6.45(4) & 2.6(1) & 0.76(3) & 0.32(5) & 1.5(1) & 0.81 (1976) \\
0606321501 & 0.10(1) & 0.64(2) & -0.52(3) & 3.59(8  & 6.48(3) & 1.2(1) & 0.77(4) & 0.33(1) & 0.9(4) & 0.74 (1976) \\
0606321601 & 0.10(1) & 0.93(7) & -0.43(1) & 6.74(8) & 6.39(7) & 3.4(3) & 0.73(5) & 0.35(2) & 3.0(3) & 0.85 (1976) \\
0606321701 & 0.09(2) & 0.40(2) & -0.49(4) & 1.61(4) & 6.59(8) & 0.9(2) & 0.78(3) & 0.34(4) & 2.2(1) & 0.73 (1976) \\
0606321801 & 0.10(1) & 0.29(1) & -0.45(2) & 1.21(6) & 6.67(4) & 0.7(1) & 0.75(2) & 0.27(5) & 0.9(2) & 0.77 (1976) \\
0606321901 & 0.10(1) & 0.23(2) & -0.42(2) & 1.09(3) & 6.59(6) & 0.6(3) & 0.78(4) & 0.21(1) & 0.8(3) & 0.78 (1976) \\
0606322001 & 0.10(1) & 0.35(1) & -0.51(1) & 2.06(5) & 6.44(4) & 1.0(1) & 0.73(5) & 0.35(2) & 1.1(2) & 0.94 (1976) \\
0606322101 & 0.10(2) & 0.27(2) & -0.53(2) & 0.71(3) & 6.70(7) & 0.8(2) & 0.78(3) & 0.18(5) & 0.7(1) & 0.83 (1976) \\
0606322201 & 0.10(1) & 0.32(1) & -0.49(1) & 1.90(4) & 6.59(3) & 1.1(3) & 0.78(4) & 0.23(2) & 1.8(2) & 0.95 (1976) \\
0606322301 & 0.11(2) & 0.33(2) & -0.54(2) & 1.09(8) & 6.58(7) & 1.0(4) & 0.76(5) & 0.20(1) & 0.9(1) & 0.74 (2157) \\
      \hline
      \end{tabular}
   \label{tab:analysis_XMM}
   \tablefoot{
$^a$ The spectral model is gabs x (BMC + Laor + redge). 
$^b$ Normalization parameters of the  BMC and Laor components are in units of $L_{35}/d_{10}^2$ erg/s/kpc$^2$ , where 
$L_{35}$ is the source luminosity in units of $10^{35}$ erg/s, $d_{10}$ is the distance to the source in units 
of 10 kpc, and the Laor and redge components are in units of $10^{-4}\times$ total photons cm$^{-2}$ s$^{-1}$ in line.
}
\end{table*}

%
%

\begin{table*}
    \caption{Best-fit parameters of spectral analysis for 2005 and 2008 with $Suzaku$ observations of 
NGC~4051 in 0.2 -- 10 keV energy range$^a$}
    \centering
    \begin{tabular}{l c c c c c c c c c c}
      \hline
      \hline
Observational& $kT_s$    & $\alpha=$  & $\log(A)$ & $N_{bmc}^b$ & $E_{line}$ & $N_{line}^b$ & $E_{c}$  & $kT_{pl}$ &  $N_{redge}^b$ &$\chi^2_{red}$ \\
ID           & (keV)   & $\Gamma-1$   &        &  (keV)      &            &              &   (keV)   &  (keV)  &    & (dof)        \\
      \hline
700004010 &  0.13(1) & 0.24(1) & 0.37(2) & 0.46(3) & 6.69(4) & 2.4(3) & 0.78(4) & 0.35(5) & 1.7(4) & 0.97 (1052) \\
703023010 &  0.29(1) & 0.81(2) & 0.48(1) & 2.48(1) & 6.42(6) & 3.6(2) & 0.75(5) & 0.32(3) & 2.1(2) & 1.04 (2207) \\
703023020 &  0.28(1) & 0.53(2) & 0.42(1) & 1.75(2) & 6.43(3) & 3.1(4) & 0.78(3) & 0.31(2) & 2.0(3) & 0.92 (1542) \\
      \hline
      \end{tabular}      
   \label{tab:analysis_Suz}
      \tablefoot{
$^a$ The spectral model is gabs x (BMC + Laor + Redge). 
$^b$ Normalization parameters of BMC and Laor components are in units of $L_{35}/d_{10}^2$ erg/s/kpc$^2$ , where 
$L_{35}$ is the source luminosity in units of $10^{35}$ erg/s, $d_{10}$ is the distance to the source in units 
of 10 kpc, and the Laor and redge components are in units of $10^{-4}\times$ total photons cm$^{-2}$ s$^{-1}$ in line; 
the parameters of the gabs model were fixed ($E_{line}$ = 0.75 keV, $\sigma$ = 0.02 keV, line depth varies from 0.0001 to 0.01).
}
\end{table*}

%
%

\begin{table*}
    \caption{Best-fit parameters of spectral analysis for 1996 and 2001 with PCA/{\it RXTE} observations of 
NGC~4051 in 3 -- 30 keV energy range$^a$}
    \centering
    \begin{tabular}{l c c c c c c c}
      \hline
      \hline
Observational& $kT_s$  & $\alpha=$   & $\log(A)$ & $N_{bmc}^b$ & $E_{line}$ & $N_{line}^b$ & $\chi^2_{red}$ \\
ID           & (keV)   & $\Gamma-1$  &            &       (keV)      &       &            &   (dof)        \\
      \hline
10301-01-04-00 & 0.11$\pm$0.02 & 0.49$\pm$0.08 & -0.02$\pm$0.01 & 1.92$\pm$0.3 & 6.59$\pm$0.08 & 0.9$\pm$0.1 & 1.02 (93) \\
10301-01-07-00 & 0.09$\pm$0.01 & 0.77$\pm$0.09 &  0.12$\pm$0.03 & 2.74$\pm$0.7 & 6.49$\pm$0.03 & 1.9$\pm$0.2 & 1.09 (62) \\
10301-01-08-00 & 0.07$\pm$0.01 & 0.85$\pm$0.09 &  0.17$\pm$0.03 & 3.32$\pm$0.2 & 6.46$\pm$0.07 & 2.1$\pm$0.2 & 0.57 (62) \\
10301-01-09-00 & 0.13$\pm$0.03 & 1.01$\pm$0.07 &  0.37$\pm$0.07 & 6.99$\pm$0.3 & 6.41$\pm$0.03 & 2.6$\pm$0.5 & 1.01 (57) \\
10301-01-22-00 & 0.05$\pm$0.01 & 0.61$\pm$0.09 & -0.01$\pm$0.01 & 0.98$\pm$0.1 & 6.67$\pm$0.09 & 0.5$\pm$0.1 & 0.68 (57) \\
10301-01-58-00 & 0.05$\pm$0.02 & 0.59$\pm$0.09 & -0.02$\pm$0.01 & 1.46$\pm$0.1 & 6.57$\pm$0.06 & 0.7$\pm$0.3 & 0.82 (57) \\
10301-01-59-00 & 0.04$\pm$0.01 & 0.77$\pm$0.05 &  0.09$\pm$0.03 & 2.79$\pm$0.1 & 6.18$\pm$0.07 & 1.7$\pm$0.2 & 1.09 (57) \\
10301-01-60-00 & 0.11$\pm$0.03 & 1.03$\pm$0.06 &  0.47$\pm$0.06 & 8.26$\pm$0.5 & 6.36$\pm$0.09 & 3.8$\pm$0.4 & 0.87 (57) \\
10301-01-61-00 & 0.06$\pm$0.01 & 0.85$\pm$0.05 &  0.19$\pm$0.04 & 4.10$\pm$0.4 & 6.43$\pm$0.08 & 1.9$\pm$0.1 & 1.24 (57) \\
20318-01-01-00 & 0.10$\pm$0.03 & 0.35$\pm$0.09 &  0.04$\pm$0.01 & 0.24$\pm$0.1 & 6.68$\pm$0.04 & 0.6$\pm$0.3 & 0.83 (57) \\
50153-01-01-00 & 0.21$\pm$0.09 & 0.57$\pm$0.09 & -0.14$\pm$0.07 & 2.97$\pm$0.3 & 6.50$\pm$0.09 & 1.6$\pm$0.2 & 0.79 (79) \\
50153-01-01-01 & 0.10$\pm$0.08 & 1.01$\pm$0.10 &  0.03$\pm$0.01 & 8.37$\pm$0.3 & 6.37$\pm$0.08 & 3.6$\pm$0.1 & 0.75 (50) \\
50153-01-01-02 & 0.10$\pm$0.08 & 0.98$\pm$0.09 &  0.97$\pm$0.14 & 6.24$\pm$0.2 & 6.43$\pm$0.09 & 2.4$\pm$0.4 & 0.85 (50) \\
50153-01-01-03 & 0.10$\pm$0.08 & 1.00$\pm$0.08 &  0.97$\pm$0.14 & 5.71$\pm$0.2 & 6.42$\pm$0.03 & 2.1$\pm$0.2 & 0.85 (50) \\
50153-01-01-05 & 0.32$\pm$0.08 & 0.34$\pm$0.08 &  0.33$\pm$0.14 & 1.28$\pm$0.2 & 6.52$\pm$0.04 & 0.9$\pm$0.3 & 0.85 (50) \\
50153-01-01-06 & 0.61$\pm$0.02 & 0.93$\pm$0.04 &  0.06$\pm$0.03 & 5.38$\pm$0.2 & 6.46$\pm$0.08 & 2.2$\pm$0.4 & 0.95 (50) \\
      \hline
      \end{tabular}      
   \label{tab:analysis_RXTE}
   \tablefoot{
$^a$ The spectral model is gabs x (BMC + Laor). 
$^b$ Normalization parameters of the BMC and the Laor components are in units of $L_{35}/d_{10}^2$ erg/s/kpc$^2$ , where 
$L_{35}$ is the source luminosity in units of $10^{35}$ erg/s, $d_{10}$ is the distance to the source in units 
of 10 kpc, and the Laor component is in units of $10^{-4}\times$ total photons cm$^{-2}$ s$^{-1}$ in line; 
the parameters of the gabs model were fixed ($E_{line}$ = 0.75 keV, $\sigma$ = 0.02 keV, line depth  varies from 0.0001 to 0.01).
}
\end{table*}

%
%
\begin{table*}
 \caption{BH masses and distances}
 \label{tab:par_scal}
 \centering 
 \begin{tabular}{lllllc}
 \hline\hline                        
      Source   & M$^a_{dyn}$ (M$_{\odot})$ & i$_{orb}^a$ (deg) & d$^b$ (kpc)  & $M_{fund.plane}$ (M$_{\odot}$) &M$_{scal}$ (M$_{\odot}$) \\
      \hline
GRO~J1655--40  &   6.3$\pm$0.3$^{(1, 2)}$ &  70$\pm$1$^{(1, 2)}$ &  3.2$\pm$0.2$^{(3)}$    &...&   ... \\
Cyg~X--1       &   6.8 -- 13.3$^{(4, 5)}$ &  35$\pm$5$^{(4, 5)}$ &  2.5$\pm$0.3$^{(4, 5)}$  &...&   7.9$\pm$1.0 \\
NGC~4051$^{(6, 7, 8, 9, 10)}$  & ... &     ...      & $\sim$9800 & $\sim$3$\times 10^6$& $\ge 6\times 10^{5}$ \\
 \hline                                             
 \end{tabular}
 \tablebib{
(1) Green et al. 2001; (2) Hjellming \& Rupen 1995; (3) Jonker \& Nelemans G. 2004. 
(4) \cite{Herrero95}; (5) \cite{Ninkov87}; 
(6) M$^c$Hardy et al. (2004); 
(7) \cite{haba08}; 
(8) Pounds \& King (2013); 
(9) \cite{Lobban11}; 
(10) \cite{Terashima09}.
}
 \end{table*}

%
%

\begin{table*} 
 \caption{Parameterizations for the reference and target sources}
 \label{tab:parametrization_scal}
 \centering 
 \begin{tabular}{lcccccc}
 \hline\hline                        
  Reference source  &       $\cal A$ &     $\cal B$     &  $\cal D$  &    $x_{tr}$      & $\beta$  &  \\
      \hline
Cyg~X--1       & 2.09$\pm$0.01 &  0.52$\pm$0.02  &  1.0 & 0.4$\pm$0.1   &   3.5$\pm$0.1  \\
GRO~J1655--40  & 2.03$\pm$0.02 &  0.45$\pm$0.03  &  1.0 & 0.07$\pm$0.02   &   1.88$\pm$0.25  \\
 \hline\hline                        
  Target source     &      $\cal A$     &    $\cal B$    &  $\cal  D$  &   $x_{tr} [\times 10^{-4}]$ & $\beta$ \\
      \hline
NGC~4051 & 2.05$\pm$0.07 & 0.61$\pm$0.08   & 1.0  &   9.56$\pm$0.07 & 0.52$\pm$0.09  \\
 \hline                                             
 \end{tabular}
 \end{table*}

\end{document}